\documentclass[pdflatex,sn-mathphys-num]{sn-jnl}

\usepackage{graphicx}
\usepackage{multirow}
\usepackage{amsmath,amssymb,amsfonts}
\usepackage{amsthm}
\usepackage{mathrsfs}
\usepackage[title]{appendix}
\usepackage{xcolor}
\usepackage{textcomp}
\usepackage{booktabs}
\usepackage{longtable}
\usepackage{array}
\usepackage{url}
\usepackage{listings}
\usepackage[expansion=false]{microtype}  % expansion disabled: TikZ EC fonts are Type 3

\usepackage{tikz}
\usepackage{pgfplots}
\pgfplotsset{compat=1.18}
\usetikzlibrary{automata,positioning,arrows.meta,shapes.geometric,
                shapes.misc,fit,backgrounds,calc}

\definecolor{snblue}{RGB}{0,70,127}
\definecolor{snlblue}{RGB}{173,210,240}
\definecolor{sngrey}{RGB}{90,90,90}
\definecolor{sngreen}{RGB}{0,128,96}
\definecolor{snorange}{RGB}{204,102,0}
\definecolor{snred}{RGB}{180,30,30}

\newcolumntype{P}[1]{>{\raggedright\arraybackslash}p{#1}}
\newcolumntype{C}[1]{>{\centering\arraybackslash}p{#1}}

\title[Closed-Loop Autonomous Software Development]{Closed-Loop Autonomous Software Development via Jira-Integrated Backlog Orchestration: A Case Study in Deterministic Control and Safety-Constrained Automation}

\author*[1]{\fnm{Elias} \sur{Calboreanu}}\email{ecalboreanu@theswiftgroup.com}
\equalcont{ORCID: 0009-0008-9194-0589}

\affil*[1]{\orgdiv{Swift North AI Lab}, \orgname{The Swift Group, LLC},
  \orgaddress{\city{Maryland}, \country{USA}}}

\abstract{This paper presents a closed-loop system for software lifecycle management
framed as a control architecture rather than a code-generation tool. The system manages a
backlog of approximately 1,602 rows across seven task families, ingests 13 structured
source documents, and executes a deterministic seven-stage pipeline implemented as seven
scheduled automation lanes. The automation stack comprises approximately 12,661 lines of
Python across 23 scripts plus 6,907 lines of versioned prompt specifications, with
checkpoint-based time budgets, 101 exception handlers, and 12 centralized lock mechanisms
implemented through four core functions and eight reusable patterns. A Jira Status
Contract provides externally observable collision locking, and a degraded-mode protocol
supports continued local operation when Jira is unavailable. Artificial-intelligence
assistance is bounded by structured context packages, configured resource caps, output
re-validation, and human review gates. A formal evaluation of the initial 152-run window yielded
100\% terminal-state success with a 95\% Clopper--Pearson interval of [97.6\%, 100\%];
the system has since accumulated more than 795 run artifacts in continuous operation.
Three rounds of adversarial code review identified 51 findings, all closed within the
study scope (48 fully remediated, 3 closed with deferred hardening), with zero false
negatives within the injected set. In an autonomous security
ticket family of 10 items, six were completed through pipeline-autonomous dispatch and
verification, two required manual remediation, and two were closed by policy decision. The
results indicate that bounded, traceable lifecycle automation is practical when autonomy is
embedded within explicit control, recovery, and audit mechanisms.}


\begin{document}

\maketitle

\section{Introduction}\label{sec:intro}
%%=============================================================

Commercial attention around AI-assisted software development has prominently emphasized
code-oriented tools such as GitHub Copilot~\cite{bib1} and Tabnine~\cite{bib2}. Yet the
software lifecycle encompasses far more than producing novel source code: it includes
backlog management, requirements consolidation, regulatory compliance, issue tracking,
change orchestration, and verification. This paper reframes automation as a control-loop
problem: given heterogeneous inputs (planning documents, issue trackers, automated
insights), produce a deterministic sequence of actions that maintains system-state
integrity while advancing work toward defined goals.

The system presented here is structured as a closed-loop feedback mechanism with explicit
state representation, policy gates, failure detection, and recovery protocols.
Artificial-intelligence involvement is bounded to a supervisory layer within deterministic
rails: no open-ended code synthesis occurs outside policy boundaries, all
machine-generated outputs are re-validated before use, and actions that affect backlog
state or code must satisfy the confidence and review policy defined in
Section~\ref{sec:arch}.

This paper investigates three research questions:

\begin{itemize}
\item \textbf{RQ1:} To what extent can a deterministic seven-stage control loop reduce
  ad-hoc human judgment for routine software backlog triage and lifecycle orchestration?

\item \textbf{RQ2:} What is the practical boundary of safe autonomous execution in a
  security-ticket family, and how should that boundary be enforced architecturally?

\item \textbf{RQ3:} What safety mechanisms are necessary to achieve a quantitatively
  defensible reliability claim without formal proof?
\end{itemize}

This paper makes eight contributions:

\begin{itemize}
\item[\textbf{C1:}] A seven-stage deterministic control-loop architecture for software
  lifecycle automation with explicit state-machine formalization
  (Section~\ref{sec:arch}).

\item[\textbf{C2:}] A formal FMEA covering 19 failure modes under the IEC~60812:2018
  severity taxonomy, with implemented detection and recovery mechanisms and partial
  empirical validation (Section~\ref{sec:safety}).

\item[\textbf{C3:}] Quantitative evidence from three rounds of adversarial
  threat-model-based testing, achieving 100\% finding resolution and zero false negatives
  within the injected set of 51 findings (Section~\ref{sec:eval}).

\item[\textbf{C4:}] A bounded-autonomy design pattern, demonstrated in one production
  deployment, applicable to lifecycle automation systems requiring verifiable control and
  operator override (Section~\ref{sec:discuss}).

\item[\textbf{C5:}] An architecture consolidation from 11 single-purpose lanes to 7
  multi-capability lanes, reducing script count by 55\% (51 to 23) and automation
  line count by 45\% (22,946 to 12,661) while preserving documented safety controls
  (Section~\ref{sec:arch}).

\item[\textbf{C6:}] A formal collision-lock protocol via Jira status transitions, designed
  to prevent concurrent agents from modifying the same ticket simultaneously
  (Section~\ref{subsec:contract}).

\item[\textbf{C7:}] A degraded-mode resilience pattern for continued local operation
  during Jira outages, featuring proactive health checking, partitioned fallback stores,
  and centralized recovery coordination (Section~\ref{subsec:degraded}).

\item[\textbf{C8:}] A specification-completeness feedback loop (Lane~7) designed to keep
  reference documents aligned with the codebase. Lane~7 is implemented and deployed but
  was introduced on March~31, 2026; empirical validation of its coverage and accuracy
  remains future work (Section~\ref{subsec:specfeedback}).
\end{itemize}

This system provides an empirical deployment of the author's MLT (MANDATE, LATTICE, TRACE)
governance stack operating together in production. Context
Engineering~\cite{bib24} formalizes the human-to-AI input layer.
MANDATE~\cite{bib25} provides the authorization framework for autonomous agentic execution.
LATTICE~\cite{bib26} defines the confidence-threshold architecture governing when the
system acts autonomously versus defers to a human. TRACE~\cite{bib27} specifies the
trusted runtime evidence requirements implemented by the audit chain. Each framework was
previously introduced separately as a working paper or preprint; this paper reports their
joint deployment in one production system, initially examined across 152 runs and now
operating continuously across seven lanes.

The remainder is organized as follows. Section~\ref{sec:problem} characterizes the
problem. Section~\ref{sec:arch} describes the architecture. Section~\ref{sec:safety}
analyzes safety and failure recovery. Section~\ref{sec:compliance} addresses compliance.
Section~\ref{sec:eval} presents quantitative evaluation. Section~\ref{sec:discuss}
discusses implications and threats. Section~\ref{sec:related} surveys related work.
Section~\ref{sec:conclusion} concludes.

%%=============================================================

\section{Problem Statement}\label{sec:problem}
%%=============================================================

Large software projects accumulate heterogeneous metadata across multiple planning
surfaces: design documents, regulatory requirements, issue trackers, code review comments,
and automated security scans~\cite{bib3}. Manually synthesizing this into coherent,
traceable action items is labor-intensive and error-prone~\cite{bib4}. Four principal
failure modes motivate this work.

\subsection{Backlog Fragmentation and Inconsistency}\label{subsec:fragmentation}

A baseline snapshot recorded 820 items in To~Do, 223 In~Progress, and 689 Done across
multiple Jira projects. These numbers have since shifted substantially following 461
AUTO-DROP bulk ticket creations and 120 ticket transitions back to To~Do during the
March~2026 consolidation period. The authoritative source of truth varied: some items
existed only in planning documents, others only in tickets, and still others were
duplicated, introducing latency and inconsistency.

\subsection{Compliance and Audit Risk}\label{subsec:compliance-risk}

The backlog included certification-related and compliance items with direct regulatory
implications. The AUTO-CERT family (29 items) tracks auto-derived certificate
requirements; the CERTREQ family (135 items in the original baseline) was subsequently
eliminated during the architectural consolidation, with its items subsumed into existing
families. Manual methods offer no verifiable evidence chain linking decision points to
outcomes.

\subsection{Distributed Source Documents}\label{subsec:distributed}

Requirements were scattered across 13 structured source documents on multiple planning
surfaces: design specifications, architecture reviews, security assessments, deployment
manifests, and operator runbooks. Without systematic parsing, deduplication, and
confidence scoring, insights were missed and effort duplicated.

\subsection{Inability to Close the Loop}\label{subsec:noloop}

No systematic mechanism existed to orchestrate issue resolution through ingestion
$\to$ canonicalization $\to$ execution $\to$ verification $\to$ publication while
maintaining audit traceability. This motivates the closed-loop design in
Section~\ref{sec:arch}.

%%=============================================================

\section{System Architecture}\label{sec:arch}
%%=============================================================

\begin{figure}[htbp]
\centering
\includegraphics[width=\textwidth]{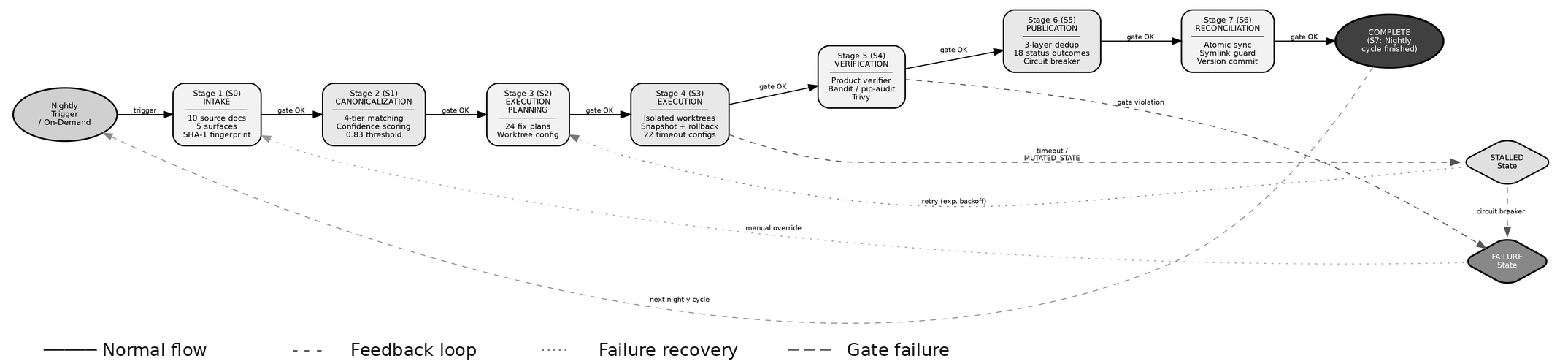}
  \caption{Seven-stage closed-loop control pipeline. Each stage gates on explicit success
    conditions, produces observable artifacts, and advances only when its exit criteria are
    satisfied. Failure at any gate transitions to the recovery cascade summarized in
    Section~\ref{subsec:recovery}.}
  \label{fig:pipeline}
\end{figure}

The system is structured as a closed-loop control architecture with seven deterministic
stages (Fig.~\ref{fig:pipeline}). Each stage produces observable artifacts, gates on
explicit success conditions, and supports failure recovery.

\subsection{Dual-Representation Model}\label{subsec:dualrep}

Two concurrent representations are maintained: a local canonical backlog (a versioned
repository, the authoritative record of system intent) and the remote Jira instance (a
shared public status surface). The backlog spans 7 families: KAN (the general-purpose Jira
backlog) approximately 1,024 items (63.9\%), AUTO-DROP 442 (27.6\%), EMU 77 (4.8\%),
AUTO-CERT 29 (1.8\%), AUTO-SEC (autonomous security) 10 (0.6\%), AUTO-OPS 10 (0.6\%), and
AUTO-TECH 10 (0.6\%), totaling approximately 1,602 rows. The CERTREQ family was eliminated
during the consolidation period, and its items were subsumed into existing families.
AUTO-DROP (bulk-created March~26--28, 2026) and AUTO-TECH represent new automated task
families.

\begin{figure}[htbp]
\centering
\includegraphics[width=0.85\textwidth]{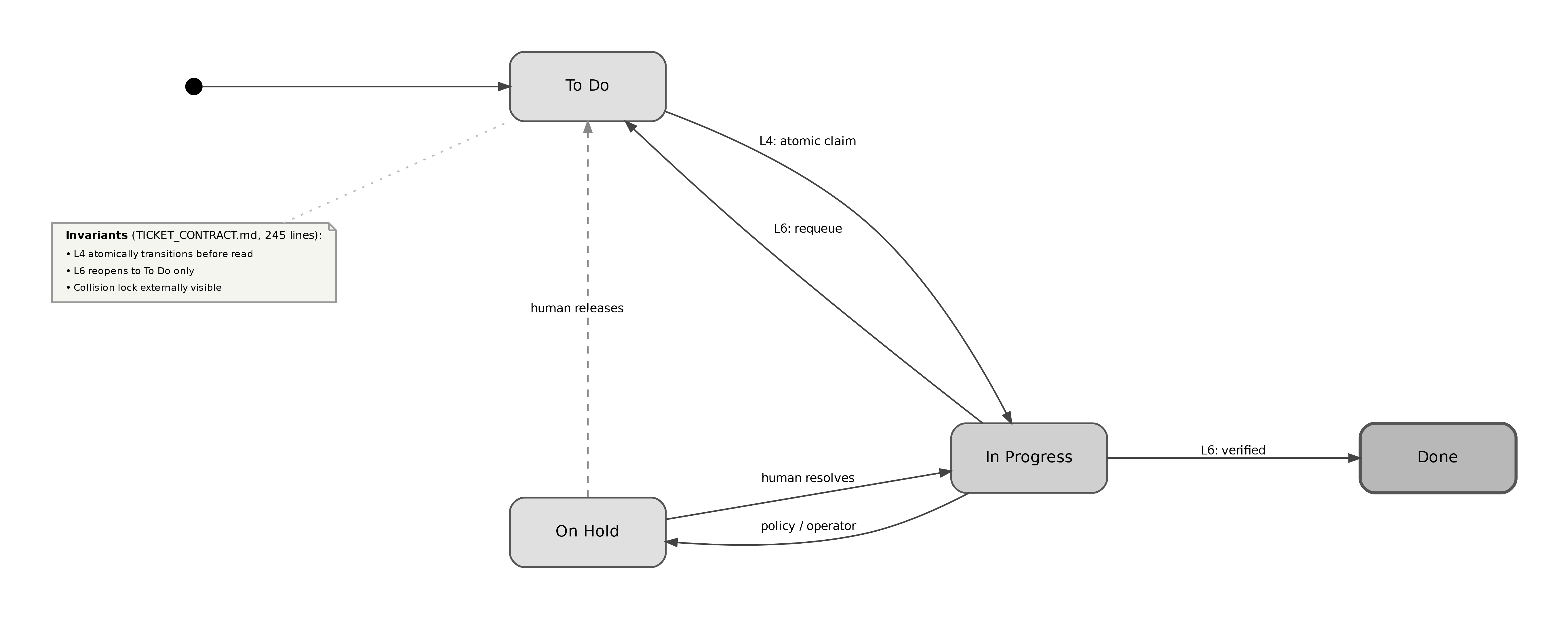}
  \caption{Jira Status Contract state machine. Four states govern individual ticket
    lifecycle: To~Do, In~Progress, On~Hold, and Done. Lane~4 atomically transitions
    To~Do to In~Progress before reading code, preventing concurrent agents from claiming
    the same ticket. Lane~6 reopens failed tickets to To~Do, never to In~Progress,
    ensuring re-entry through the grooming pipeline. Safety invariants are codified in
    \texttt{TICKET\_CONTRACT.md} (245~lines).}
  \label{fig:jira_sm}
\end{figure}

\subsection{Seven-Stage Control Loop}\label{subsec:controlloop}

The seven conceptual stages are implemented as 7 consolidated automation lanes, each
operating on a scheduled cadence rather than as a monolithic nightly batch. Lane~1
(Backlog Sync and Intake, hourly) reads source documents and synchronizes with Jira.
Lane~2 (Codebase Auditor, every 2~hours) scans the live codebase in read-only mode,
writing findings to \texttt{lane\_02\_untracked\_findings.json} for Lane~3 ingestion.
Lane~2 has no direct Jira write authority during normal operation; in degraded mode it can
emit operator-facing fallback notes, but ticket creation remains downstream. Lane~3
(Backlog Groomer, every 4~hours) deduplicates, maps items with confidence scoring, and
selects fix strategies. Lane~4 (AI Code Fixer, hourly) applies changes in isolated
worktrees with rollback. Lane~5 (Ops Intelligence, every 2~hours) monitors lane health,
Jira connectivity, and ticket-status distribution. Lane~6 (Quality Gate, every 2~hours)
runs product and security verifiers and posts results to Jira with three-layer
deduplication. Lane~7 (Spec Completeness Auditor, every 4~hours) checks whether
specification documents reflect the deployed codebase.

Rather than processing the entire backlog in a single nightly sweep, each lane
continuously processes its domain at its own cadence. Lanes~3, 4, and~6 implement an
Adaptive Worker Pool for parallel sub-agent execution within each lane. The worker count
per lane is computed as:
\begin{equation*}
  \mathit{workers} = \min\!\Bigl(
    \mathit{actionable},\;
    \mathit{MAX\_WORKERS},\;
    \bigl\lfloor \mathit{t\_remain} / \mathit{t\_avg} \bigr\rfloor
  \Bigr)
\end{equation*}
where $\mathit{MAX\_WORKERS}$ is 3 for Lanes~3 and~4 (with worktree isolation) and 2 for
Lane~6. Lane~4 additionally operates in a dual-mode configuration: hourly standard runs
with a 45-minute time budget for routine fixes, and a nightly deep sweep at 1:30~AM with a
120-minute budget targeting \texttt{ai\_assisted} category tickets with full test-suite
execution and worktree isolation.

Lane~3 outputs a structured \texttt{fix\_queue} consumed by Lane~4, with schema fields
including \texttt{priority\_score} (numeric), \texttt{priority\_type}
(e.g.,~\texttt{dead\_ui\_button}), \texttt{relevant\_paths}, \texttt{area}, and
\texttt{confidence}. Lane~4 processes this queue sorted by \texttt{priority\_score}
descending.

Mapping confidence across the backlog: items scoring $s \geq 0.83$ may be eligible for
autonomous merge within pre-approved task families; items scoring $0.50 \leq s < 0.83$
are routed to human-in-the-loop review; items scoring $s < 0.50$ are routed for
re-ingest. All backlog rows have confirmed Jira key assignments with bidirectional
synchronization active.

\subsection{Formal State Machine}\label{subsec:fsm}

\begin{figure}[htbp]
\centering
\includegraphics[width=\textwidth]{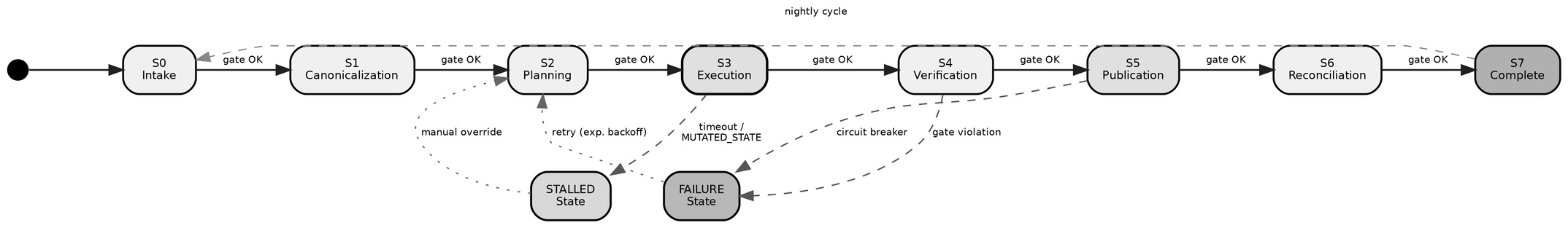}
  \caption{Formal state machine. States S0--S6 correspond to pipeline stages and S7 is
    the terminal Complete state. Failure and Stalled states enable deterministic recovery
    on the next scheduled cycle without a full restart. No hidden state exists in any
    transition.}
  \label{fig:fsm}
\end{figure}

The system models state transitions via a finite-state machine (Fig.~\ref{fig:fsm}) with
states S0--S6 corresponding to the seven stages, plus Failure and Stalled states.
Transitions are deterministic, dependent only on observable state and input conditions; no
hidden state exists. Failure states are entered on gate violations, timeout expiration, or
uncaught exceptions. Stalled states enable next-cycle recovery via retry or manual
override without a full restart.

In addition to the pipeline stage states, a secondary state machine governs individual
ticket lifecycle via the Jira Status Contract (Fig.~\ref{fig:jira_sm};
Section~\ref{subsec:contract}): To~Do (ready for processing), In~Progress (agent collision
lock), On~Hold (human-in-the-loop required), and Done (work completed and verified). The
Jira Status Contract is codified in a dedicated specification document
(\texttt{TICKET\_CONTRACT.md}, 245~lines) serving as the single source of truth for
status-transition IDs, creation authority by lane, mandatory labels, worker-pool
configuration, MCP rate limits, issue-type rules, and safety limits.

\subsection{Intake Pipeline}\label{subsec:intake}

Before automated intake, the operator executes the Reviewer stage of the Context
Engineering methodology~\cite{bib24}: a structured human-AI pass that reads the raw source
documents, identifies requirements and gaps, and produces the prioritized, role-annotated
context package that Stage~1 consumes. This human-side step ensures the pipeline receives
verified, authority-ranked inputs rather than uninterpreted raw files. The intake pipeline
(908~LOC) then processes those outputs, supporting Markdown, plain text, and DOCX formats,
applying SHA-1 fingerprinting to detect inter-run changes, and enforcing a seven-field
quality standard: unique identifier, title, description, source reference, tags, priority,
and owner. Fingerprinting enables idempotent intake across repeated reads of unchanged
documents.

\subsection{Canonicalization Engine}\label{subsec:canon}

The canonicalization engine (2,841~LOC) consolidates disparate inputs via a four-tier
matching strategy: (1)~exact tag matching, (2)~key matching, (3)~summary similarity via
weighted sequence-matching (SequenceMatcher ratio plus Jaccard token overlap), and
(4)~fuzzy text matching using Levenshtein distance. The system invokes up to 6 MCP Jira
tools across lanes (\texttt{searchJiraIssuesUsingJql}, \texttt{getJiraIssue},
\texttt{createJiraIssue}, \texttt{editJiraIssue}, \texttt{addCommentToJiraIssue},
\texttt{transitionJiraIssue}), replacing the direct HTTP API calls used in the initial
deployment.

The operational similarity score $s$ is defined as:
\begin{equation}
  s = 0.6 \cdot \mathit{SM} + 0.4 \cdot \mathit{Jac}
\label{eq:similarity}
\end{equation}
where $\mathit{SM}$ is the SequenceMatcher ratio and $\mathit{Jac}$ is the Jaccard token
overlap. Equation~(\ref{eq:tfidf}) presents the TF-IDF cosine-similarity formulation that
motivates the key-plus-summary matching component; $s$ approximates this measure for
practical computation:
\begin{equation}
  \mathrm{sim}(d_i, d_j)
    = \frac{\displaystyle\sum_{k} w_{ik} \cdot w_{jk}}%
           {\displaystyle\sqrt{\sum_{k} w_{ik}^{2}} \cdot \sqrt{\sum_{k} w_{jk}^{2}}}
\label{eq:tfidf}
\end{equation}
where $w_{ik} = \mathrm{tf}(t_k, d_i) \cdot \mathrm{idf}(t_k)$.

Items are assigned to confidence tiers based on the computed similarity score:
\begin{equation}
  \mathrm{tier}(s) =
  \begin{cases}
    \text{autonomous action}        & s \geq 0.83 \\
    \text{human-in-the-loop review} & 0.50 \leq s < 0.83 \\
    \text{halt and re-ingest}       & s < 0.50
  \end{cases}
\label{eq:tiers}
\end{equation}

The 0.83 threshold was set as a configurable parameter (default 0.83) and refined through
iterative operator review of mapping outcomes across successive consolidation runs. Items
below 0.50 are treated as new issues requiring full intake processing. The three-tier
confidence structure directly instantiates the LATTICE confidence threshold
architecture~\cite{bib26}, which specifies that systems operating above a high-confidence
threshold may act autonomously, systems in the mid-range require human verification before
proceeding, and systems below the lower bound must halt rather than propagate low-confidence
decisions.

\subsection{Execution Engine}\label{subsec:execution}

Execution occurs in isolated worktrees (clean checkouts with no shared state).
Approximately 120 mapped fix plans are maintained, spanning apply actions (code
modification, refactoring, policy updates) and verify/manual actions. Each execution is
subject to a checkpoint-based time budget: 45~minutes for standard hourly runs, 120~minutes
for nightly deep sweeps, with a 10-minute progress checkpoint that extends to 20~minutes if
forward progress is detected before hard stop. On expiration, SIGKILL is issued after a
30-second USR1 warning signal.

\subsection{Verification Layer}\label{subsec:verification}

Two post-execution verifiers run: (1)~the product verifier (5,142~LOC) ensures
compilation, unit tests, and integration constraints pass; (2)~the security verifier
(225~LOC) runs \texttt{security-audit.sh} (wrapping \texttt{npm~audit} for dependency
vulnerability scanning), \texttt{sast-scan.sh} (wrapping eslint~\cite{bib5} for static
analysis and lint auto-fix), and \texttt{verify\_backlog\_security\_ticket.py}
(pattern-match verification of security surface hardening).
\texttt{FAIL\_VERIFY\_MUTATED\_STATE} detects unexpected worktree mutations during
verify-only plans, triggering mandatory manual review.

\subsection{Publication Pipeline}\label{subsec:publication}

The publication pipeline (1,839~LOC) posts results to Jira via distinct status outcomes
including explicit transition paths: Done (ID~41), To~Do (ID~11, for requeue), and
On~Hold (ID~31, for human-in-the-loop review). Lane~6 (Quality Gate) reopens failed
tickets to To~Do rather than In~Progress, ensuring they re-enter the grooming pipeline via
Lane~3 without bypassing the collision lock. Three deduplication layers prevent duplicate
postings: local receipt log, Jira comment history, and per-run cache. A circuit breaker
prevents single malformed items from blocking the run. Zero duplicate posts were recorded
across the evaluation period.

\subsection{Reconciliation}\label{subsec:reconciliation}

Reconciliation atomically synchronizes Jira results to the canonical backlog. A symlink
boundary check prevents path-traversal vulnerabilities. Stale files are cleaned up and the
canonical backlog is re-versioned with a commit encoding the run~ID, timestamp, and change
summary.

\subsection{Bounded AI Supervisory Layer}\label{subsec:ai}

Claude (Anthropic)~\cite{bib6}, orchestrated via Cowork with MCP Jira tools, is embedded
as a bounded supervisory layer. The system shifted from GPT-5.4 (OpenAI Codex) to Claude
between the initial deployment and the current v3 architecture; the bounded-autonomy design
is model-agnostic by construction, so safety properties are enforced by architectural rails
rather than model-specific behavior. Specific decision points (requirements summarization,
test-case outlining, policy drafting) are delegated within explicit resource and review
caps: lane-specific time budgets, environment-variable-driven retry limits (default~3 for
Jira API calls), ticket caps, and diff-scrutiny tiers (under 50~lines: standard review;
50--200~lines: extended verification; over 200~lines: mandatory human review). All
AI-generated outputs are subject to format re-validation.

Configuration is managed through versioned \texttt{PROMPT.md} files (6,907 total lines
across 7~lanes, each with a semantic version header and changelog), YAML configuration
files per lane, and environment variables. Prompts are version-controlled alongside the
codebase and structured according to the context-package formalism described by
Calboreanu~\cite{bib24}: each AI call carries an explicit Authority file (the current
lane's gate criteria), a Source file (the item under evaluation), and a Conductor prompt
(the action directive). Current PROMPT versions are: Lane~1 v2.3.0, Lane~2 v2.4.0,
Lane~3 v2.7.0, Lane~4 v3.7.0, Lane~5 v2.6.0, Lane~6 v2.6.0, Lane~7 v1.1.0.

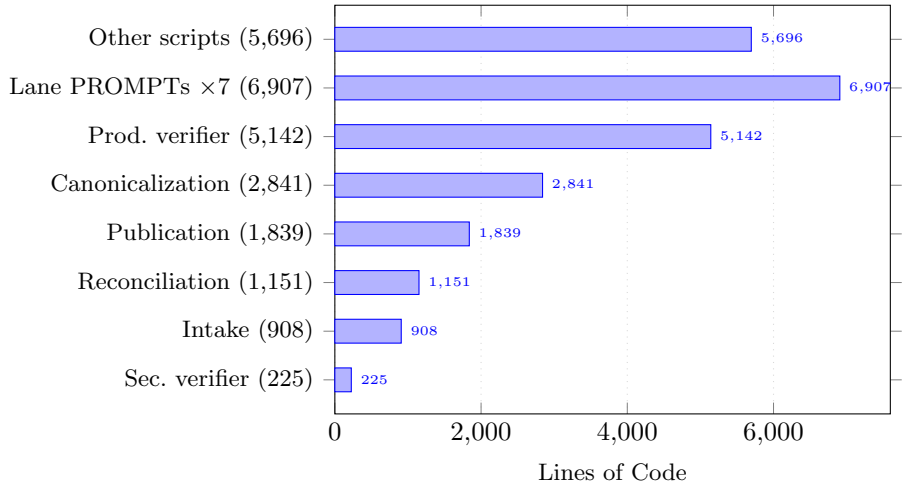
\begin{figure}[htbp]
\centering
\begin{tikzpicture}
\begin{axis}[
  xbar, xmin=0,
  width=0.68\linewidth, height=7.0cm,
  bar width=9pt,
  xlabel={Lines of Code},
  xlabel style={font=\small},
  ytick=data,
  symbolic y coords={SecVerif,Intake,Reconcile,Publish,Canon,ProdVerif,PROMPTs,Scripts},
  yticklabels={
    {Sec.\ verifier (225)},
    {Intake (908)},
    {Reconciliation (1{,}151)},
    {Publication (1{,}839)},
    {Canonicalization (2{,}841)},
    {Prod.\ verifier (5{,}142)},
    {Lane PROMPTs $\times$7 (6{,}907)},
    {Other scripts (5{,}696)}},
  yticklabel style={font=\small},
  nodes near coords,
  nodes near coords style={font=\tiny,anchor=west},
  xmajorgrids=true, grid style={dotted,gray!30},
  enlarge y limits=0.10,
  every axis plot/.append style={fill=snlblue,draw=snblue},
]
\addplot coordinates{
  (225,SecVerif)(908,Intake)(1151,Reconcile)
  (1839,Publish)(2841,Canon)(5142,ProdVerif)
  (6907,PROMPTs)(5696,Scripts)};
\end{axis}
\end{tikzpicture}
  \caption{Automation infrastructure component sizes in lines of code. The v3 architecture
    comprises 12,661 script lines plus 6,907 prompt lines across 7 lanes.}
  \label{fig:components}
\end{figure}

\subsection{Jira Status Contract and Collision Lock}\label{subsec:contract}

The Jira Status Contract externalizes ticket ownership and status transitions so that
mutually exclusive work claims are visible to every lane and every human operator. Lane~4
must atomically transition a ticket from To~Do to In~Progress before reading code; if the
transition fails, the worker abandons the claim and selects a different item. Lane~6 is the
only lane that reopens failed tickets, and it does so by returning them to To~Do so that
they re-enter the grooming pipeline rather than bypassing it. This contract is intentionally
narrower than a general workflow engine: it exists to make concurrency, provenance, and
re-entry rules explicit.

\subsection{Degraded-Mode Coordination}\label{subsec:degraded}

When Jira is unavailable, the system switches from connected publication to local
continuity. Lane~1 writes outbound sync intents to
\texttt{jira\_write\_outbox.json}. Lanes~3, 4, and~6 queue publication intents to
\texttt{mcp\_replay\_needed.json}. Lanes~2 and~7 may emit operator-facing fallback notes
to \texttt{jira\_fallback\_queue.md}, but they do not acquire direct ticket-creation
authority in degraded mode. Lane~5 acts as the recovery coordinator: it detects restored
connectivity, replays queued write intents in bounded batches, clears degraded status, and
reports the recovery outcome.

\subsection{Specification-Completeness Feedback}\label{subsec:specfeedback}

Lane~7 closes the loop between implementation and reference documents by checking whether
the specification set still reflects the deployed codebase. It compares codebase
observations against the structured source-document corpus, flags missing or stale
statements, and prepares proposed updates for review. Because Lane~7 was introduced on
March~31, 2026, the paper treats it as an implemented mechanism whose empirical evaluation
remains future work rather than as a completed validation result.

%%=============================================================

\section{Safety and Failure Recovery}\label{sec:safety}
%%=============================================================

Given that the system makes autonomous lifecycle decisions, safety is a primary design
constraint. The system adopts a defense-in-depth strategy~\cite{bib7} combining multiple
independent mechanisms.

\subsection{Bounded Autonomy Model}\label{subsec:bounded}

All decisions occur within explicitly defined task families, subject to rate limits,
resource caps, and policy gates. The authorization model implements the MANDATE
framework~\cite{bib25}, which answers the question: given a proposed autonomous action and
the current system state, is this agent authorized to execute? The approximately 120 fix
plans constitute the MANDATE task taxonomy for this deployment: each plan specifies the
action type, the preconditions that must hold, the verification steps that must be passed,
and the human escalation path if they are not. Additionally, the collision-lock protocol
(Section~\ref{subsec:contract}) adds a pre-condition to every fix plan: the ticket must be
in To~Do status and successfully transitioned to In~Progress before work begins. The
autonomous security (AUTO-SEC) family (10 items, 0.6\% of the backlog) represents the
maximum authorized autonomous scope under the current policy configuration; even these are
limited to pre-approved types and subject to identical verification and publication gating
as manually authored items. Any expansion of the authorized scope requires an explicit
policy change, a fix plan update, and re-validation against the FMEA.

\subsection{Failure Mode and Effects Analysis}\label{subsec:fmea}

Table~\ref{tab:fmea} presents a formal FMEA under IEC~60812:2018~\cite{bib8}, expanded
from 12 to 19 failure modes in the v3 architecture. New failure modes include Jira
connectivity loss during mid-fix (mitigated by the degraded-mode protocol and local lock
fallback), multi-agent collision on the same ticket (mitigated by atomic Jira status
transition as lock), stale In~Progress tickets from crashed agents (mitigated by Lane~5
monitoring with 4-hour WARNING and 10-hour CRITICAL thresholds), aging On~Hold tickets not
actioned by humans (mitigated by Lane~5 48-hour alerting), intra-lane worker collision
(mitigated by per-worker ticket claim via Adaptive Worker Pool queue), worker pool resource
exhaustion (mitigated by time-budget checkpoint model and hard stop), and time-budget
exhaustion mid-ticket (mitigated by worktree discard and ticket requeue to To~Do). Severity
uses the IEC four-tier scale: catastrophic~(IV), critical~(III), marginal~(II), and
negligible~(I). Ratings were assigned by a two-person review panel through a structured
walkthrough against documented system requirements.

The architectural separation between execution (Stage~4) and verification (Stage~5)
reflects Rule~6 of the context engineering methodology~\cite{bib24}: ``the executor cannot
be the auditor,'' applied at the machine level. The \texttt{FAIL\_VERIFY\_MUTATED\_STATE}
check is the enforcement mechanism for this rule: it is a machine-level tripwire that
automatically catches any violation of the execution-verification boundary.

\subsection{Lock and Concurrency Control}\label{subsec:locks}

Twelve centralized lock mechanisms protect critical sections. These mechanisms are
implemented through 4~core functions in \texttt{shared/lock\_handler.py}
(\texttt{acquire\_lock}, \texttt{release\_lock}, \texttt{is\_lock\_stale},
\texttt{clear\_stale\_lock}) and 8~reusable lock patterns across lane implementations;
7~per-lane lock files instantiate those mechanisms with configurable TTLs (Lane~1
TTL~=~2~h, Lane~7 TTL~=~4~h). Additionally, \texttt{jira\_connectivity.json} serves as a
distributed coordination file for degraded-mode detection, and
\texttt{.jira\_ticket\_locks.json} provides local collision protection when Jira is
unavailable. Lock semantics include PID verification, clock-skew clamping, and race-safe
re-acquire, providing fine-grained concurrency control without global serialization. The
consolidation from 58 distributed lock functions to 12 centralized mechanisms reflects
architectural simplification during the 11-to-7 lane consolidation.

\subsection{Timeout and Process Management}\label{subsec:timeout}

Across the lane fleet, checkpoint-based time budgets are configured through a checkpoint
model. Lane~4 uses a 45-minute standard time budget (120~minutes for nightly deep sweeps)
with a 10-minute progress checkpoint that extends to 20~minutes if forward progress is
detected. Lane~2 enforces a 30-minute hard stop, Lane~7 a 45-minute hard stop, and other
lanes apply lane-specific limits consistent with their schedules and lock TTLs (for
example, Lane~1 TTL~=~2~h; Lane~7 TTL~=~4~h). On hard-stop expiration, a USR1 warning is
followed by SIGKILL after 30~seconds.

\subsection{Failure Recovery Cascade}\label{subsec:recovery}

On failure, six mechanisms engage in sequence: (1)~worktree discard and immediate retry;
(2)~receipt-log replay from last checkpoint; (3)~exponential back-off with jitter, defined
by:
\begin{equation}
  t_n = \min\!\bigl(t_{\max},\; t_0 \cdot 2^{n}\bigr)
        \cdot \bigl(1 + \varepsilon\bigr), \quad
        \varepsilon \sim \mathcal{U}(-0.10,\, +0.10);
\label{eq:backoff}
\end{equation}
(4)~circuit breaker escalation after three consecutive failures; (5)~crash handler
transition to Stalled state, preserving all evidence for post-mortem; (6)~queue drain
recovery, in which Lane~5 replays queued Jira writes on connectivity recovery, processing
up to 20~MCP calls per drain run in oldest-first order.

\subsection{Offline Continuity}\label{subsec:offline}

The canonical backlog serves as an operational cache enabling continued planning and
verification when Jira is unreachable. All 7~lanes implement a proactive degraded-mode
protocol: before any MCP Jira call, each lane reads \texttt{jira\_connectivity.json}; if
the status is DEGRADED, all MCP calls are skipped immediately. The first lane to encounter
a Jira failure writes the degraded-status file and switches to offline mode. Lanes queue
pending publication intents to three partitioned fallback stores:
\texttt{mcp\_replay\_needed.json} (Lanes~3, 4, and~6),
\texttt{jira\_fallback\_queue.md} (operator-facing review notes from Lanes~2 and~7), and
\texttt{jira\_write\_outbox.json} (Lane~1). Lane~2 does not acquire direct Jira
ticket-creation authority in degraded mode; its fallback entries remain informational
artifacts for later reconciliation. Lane~4 falls back to a local lock file
(\texttt{.jira\_ticket\_locks.json}) when Jira is unavailable, maintaining collision
protection even during outages. Lane~5 acts as the Jira recovery coordinator: on
connectivity recovery, it replays queued write intents (max 20~MCP calls per drain run),
clears the degraded status, and reports a recovery summary.

\subsection{Health Watchdog}\label{subsec:watchdog}

Lane~5 (Ops Intelligence) serves as the unified health monitor for all 7~automation lanes,
replacing the original three-lane watchdog. Lane~5 runs every 2~hours and checks: lock
file staleness per lane, report freshness, scheduled task liveness, Jira ticket status
distribution (stale In~Progress tickets exceeding 4~hours trigger a WARNING; exceeding
10~hours triggers a CRITICAL alert; aging On~Hold tickets exceeding 48~hours are flagged),
Jira connectivity status, and queue drain status for degraded-mode recovery. Lane~5
produces a comprehensive daily digest at 7:15~AM. Threshold breaches trigger alerts and
escalate to manual review, catching process hangs and silent failures across the entire
lane fleet.

%%=============================================================

\section{Compliance and Auditability}\label{sec:compliance}
%%=============================================================

For safety-critical systems, auditability is a core architectural requirement~\cite{bib9}.
The system produces a complete, deterministic, externally verifiable evidence chain
designed to satisfy both internal quality assurance and external regulatory scrutiny.

\subsection{Separation of Concerns}\label{subsec:separation}

The canonical backlog is the authoritative record of system intent; Jira is the shared
public representation. If Jira is corrupted or its state diverges from the canonical
source, the system recovers by re-syncing from the canonical backlog rather than treating
Jira as ground truth. This separation ensures that no single external dependency can
compromise the authoritative record. The canonical backlog is maintained as a structured
local store with schema validation; Jira serves as the collaboration and visibility layer.
Changes flow unidirectionally from the canonical source to Jira during publication
(Stage~6) and are reconciled bidirectionally during the reconciliation stage (Stage~7).

\subsection{Evidence Chain}\label{subsec:evidence}

Each scheduled cycle produces evidence artifacts: structured logs recording every decision
point, Jira receipts confirming each status transition, and summary reports aggregating
per-lane metrics. During the initial 152-run evaluation window, related cycles were
sometimes grouped under shared directory prefixes; the audited sample therefore comprised
167~logs and 38~receipts across 73~run directories representing 152~runs. Since the
transition to continuous operation, the evidence chain has scaled proportionally with more
than 795~run artifacts accumulated as of March~31, 2026.

The structure and content of these records implement the TRACE trusted runtime evidence
specification~\cite{bib27}, which requires that every autonomous action produce a
hash-linked, tamper-evident record traceable to the authorizing policy. Each evidence
artifact includes the run identifier, lane identifier, timestamp, input file checksums,
output file checksums, exception counts, and a terminal status indicator. The hash-linking
ensures that retroactive modification of any single record is detectable through chain
validation.

Calboreanu~\cite{bib24} observes that audit trails form organically when pipeline
execution is documented (Finding~10), documenting a 1,345-line trail across 15 or more
human-AI sessions on the same project. The closed-loop architecture scales this property
to automated machine operation: the 167~evidence logs produced during the evaluation
window are system-enforced outputs, not discretionary documentation overhead.

\subsection{Change Traceability}\label{subsec:traceability}

Every Jira status transition records the originating run~ID, a UTC timestamp, and a
permalink to the corresponding evidence log. This three-element record enables any
reviewer to trace a ticket's state change back to the specific automation run that caused
it, the evidence log that documents the decision, and the policy configuration that
authorized it. The traceability chain is consistent with NIST SP~800-53 AU-12 (Audit
Record Generation), which requires that information systems generate audit records for
defined auditable events, and AU-3 (Audit Record Content), which specifies the minimum
content for audit records including event type, timestamp, source, and
outcome~\cite{bib10}.

During the evaluation window, the 38~Jira receipts across 30~unique issues each contain
the run ID, transition ID (one of four: 11~To~Do, 21~In~Progress, 31~On~Hold, 41~Done),
and the evidence log permalink. In the subsequent continuous-operation period, 120
additional tickets were transitioned with the same traceability structure.

\subsection{Conservative Status Alignment}\label{subsec:conservative}

When item status is ambiguous---for example, when an AI-generated confidence score falls
between the autonomous and halt thresholds---the system defaults to the more conservative
(less autonomous) state. AI-generated summaries below the confidence threshold are marked
for manual review rather than auto-transitioned. This conservative-default principle
applies at every decision point in the pipeline: intake items with ambiguous family
classification are routed to human review, canonicalization results below the 0.83
confidence threshold require human approval before execution, and publication actions that
would transition a ticket to Done are gated on verification-stage approval.

The conservative-default pattern ensures that automation errors fail toward human
oversight rather than toward unsupervised action, consistent with the bounded-autonomy
model described in Section~\ref{subsec:bounded}.

%%=============================================================

\section{Quantitative Evaluation}\label{sec:eval}
%%=============================================================

\subsection{System Scale}\label{subsec:scale}

The automation infrastructure comprises approximately 12,661~lines across 23~Python scripts
plus 6,907~lines of versioned PROMPT specifications (6,907~lines across 7~lane PROMPTs
plus 245~lines of \texttt{TICKET\_CONTRACT.md}), with per-lane environment variables
(including run-time budgets, ticket caps, and MCP rate limits), checkpoint-based time
budgets, 101~exception handlers, and 12~centralized lock mechanisms implemented through
4~core lock-handler functions and 8~reusable patterns. Seven per-lane lock files
instantiate those mechanisms at run time.

The reduction from the v2 metrics (22,946~lines across 51~scripts, 111~environment
variables, 58~lock functions) reflects the architectural consolidation from 11
single-purpose lanes to 7 multi-capability lanes; complexity was centralized rather than
eliminated. This governs a 1,290,158-line application codebase across 15,840 source files:
639,253~lines of JavaScript, JSX, TypeScript, and CSS across 13,154 frontend files;
609,107~lines of Python across 1,838 backend, test, and infrastructure files; 10,654~lines
of shell scripting across 61 build and operations files; and 31,144~lines of configuration
and data files (JSON, YAML, TOML, Dockerfile) across 787~files. Infrastructure density is:
\begin{equation}
  \rho = \frac{\mathrm{LOC}_{\text{automation}}}{\mathrm{LOC}_{\text{application}}}
       = \frac{12{,}661 + 6{,}907}{1{,}290{,}158} \approx 1.52\%
\label{eq:density}
\end{equation}

The reduction from 1.78\% in the v2 architecture reflects lane consolidation rather than a
change in the managed application baseline. Table~\ref{tab:scale} presents complete scale
metrics.

\subsection{Operational History and Reliability}\label{subsec:reliability}

Table~\ref{tab:reliability} summarizes the initial 152-run evaluation window across three
run types. All 152~runs achieved 100\% terminal-state success (no run entered an
unrecoverable state). Since the initial evaluation, the system has transitioned from batch
nightly runs to continuous per-lane scheduled execution: Lane~1 (hourly), Lane~2 (every
2~hours), Lane~3 (every 4~hours), Lane~4 (hourly), Lane~5 (every 2~hours), Lane~6 (every
2~hours), and Lane~7 (every 4~hours). As of March~31, 2026, more than 795~run artifacts
exist in the reporting directory.

Lane~5 (v2.6.0) includes built-in evaluation-window tracking: it reads
\texttt{evaluation\_window.json}, tracks cumulative metrics (autonomous fixes, regression
catches, human-review escalations, end-to-end resolutions) during defined windows, and
includes evaluation-window progress in operational reports, providing self-referential
instrumentation for empirical methodology. The original 79.3\% nightly DEGRADED rate
reflected informational dependency-audit alerts; all DEGRADED runs completed successfully.
DEGRADED denotes informational alerts present, not partially failed.

The Clopper--Pearson exact confidence interval~\cite{bib11} for a binomial proportion is:
\begin{equation}
  [\hat{\theta}_L,\, \hat{\theta}_U] =
  \left[\,
    B\!\left(\tfrac{\alpha}{2};\; k,\; n-k+1\right),\;
    B\!\left(1-\tfrac{\alpha}{2};\; k+1,\; n-k\right)
  \right]
\label{eq:cp}
\end{equation}
where $B(\cdot\,;\cdot,\cdot)$ denotes the beta distribution quantile function (by
convention, when $k = n$ the upper bound evaluates to 1.000). Applied to 152~consecutive
successful runs ($k = n = 152$), the interval is:
\[
  \hat{\theta}_L = B(0.025;\; 152,\; 1) \approx 0.976,\qquad
  \hat{\theta}_U = 1.000
\]
yielding a 95\% CI of $[97.6\%,\; 100\%]$.

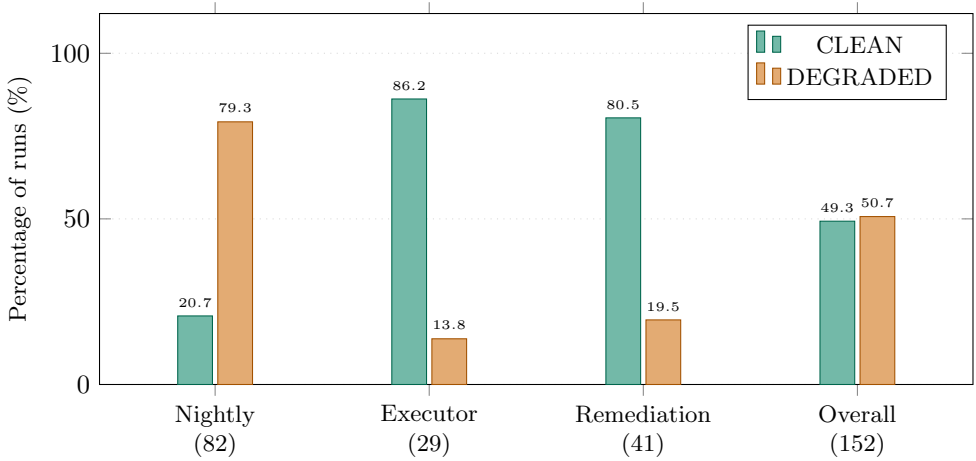
\begin{figure}[htbp]
\centering
\begin{tikzpicture}
\begin{axis}[
  ybar, bar width=13pt,
  width=\linewidth, height=6.5cm,
  ymin=0, ymax=112,
  ylabel={Percentage of runs (\%)},
  ylabel style={font=\small},
  symbolic x coords={Nightly,Executor,Remediation,Overall},
  xtick=data,
  xticklabels={{Nightly\\(82)},{Executor\\(29)},{Remediation\\(41)},{Overall\\(152)}},
  xticklabel style={font=\small,align=center},
  legend style={font=\small,at={(0.97,0.97)},anchor=north east},
  nodes near coords,
  nodes near coords style={font=\tiny},
  ymajorgrids=true, grid style={dotted,gray!30},
  enlarge x limits=0.18,
  every node near coord/.append style={/pgf/number format/.cd,fixed,precision=1},
]
\addplot[fill=sngreen!55,draw=sngreen!80!black]
  coordinates{(Nightly,20.7)(Executor,86.2)(Remediation,80.5)(Overall,49.3)};
\addplot[fill=snorange!55,draw=snorange!80!black]
  coordinates{(Nightly,79.3)(Executor,13.8)(Remediation,19.5)(Overall,50.7)};
\legend{CLEAN,DEGRADED}
\node[font=\tiny,align=center,fill=sngreen!12,draw=sngreen,
      rounded corners=2pt,inner sep=3pt] at (axis cs:Overall,96)
  {100\% success\\all 152 runs};
\end{axis}
\end{tikzpicture}
  \caption{CLEAN vs.\ DEGRADED run status by type for the initial 152-run evaluation
    window. The high nightly DEGRADED rate reflects informational dependency-audit alerts
    that do not prevent successful run completion. All 152~runs achieved terminal-state
    success. Post-evaluation, the system transitioned to continuous per-lane execution with
    more than 795 total run artifacts as of March~31, 2026.}
  \label{fig:cleandegraded}
\end{figure}

To contextualize these results against published benchmarks, the DORA Accelerate State of
DevOps report identifies four performance tiers based on deployment frequency, lead time,
change failure rate, and recovery time~\cite{bib34}. The system's continuous per-lane
cadences (hourly to every 4~hours) and sub-120-minute cycle times align with DORA's elite
tier, which reports on-demand deployment with sub-day lead times, achieved by only 19\% of
surveyed organizations~\cite{bib35}. The observed 100\% terminal-state success rate (95\% CI
[97.6\%, 100\%]) within the 152-run evaluation window is noted alongside the U.S.\ industry
average defect removal efficiency of approximately 85\%, with best-in-class organizations
achieving above 99\%~\cite{bib36}; however, the comparison is indirect given different
measurement contexts.
For security remediation, industry mean time to remediate for critical vulnerabilities
averages 60 to 150~days~\cite{bib37}; the AUTO-SEC family resolved 6 of 10~tickets
autonomously within the evaluation window. These comparisons are necessarily qualitative:
the system operates as a lifecycle orchestrator rather than a traditional CI/CD pipeline,
and the evaluation reflects a single deployment rather than a controlled experiment.
Nevertheless, the benchmarks provide useful context.

\subsection{Adversarial Code Review}\label{subsec:adversarial}

\begin{figure}[htbp]
\centering
\includegraphics[width=0.75\textwidth]{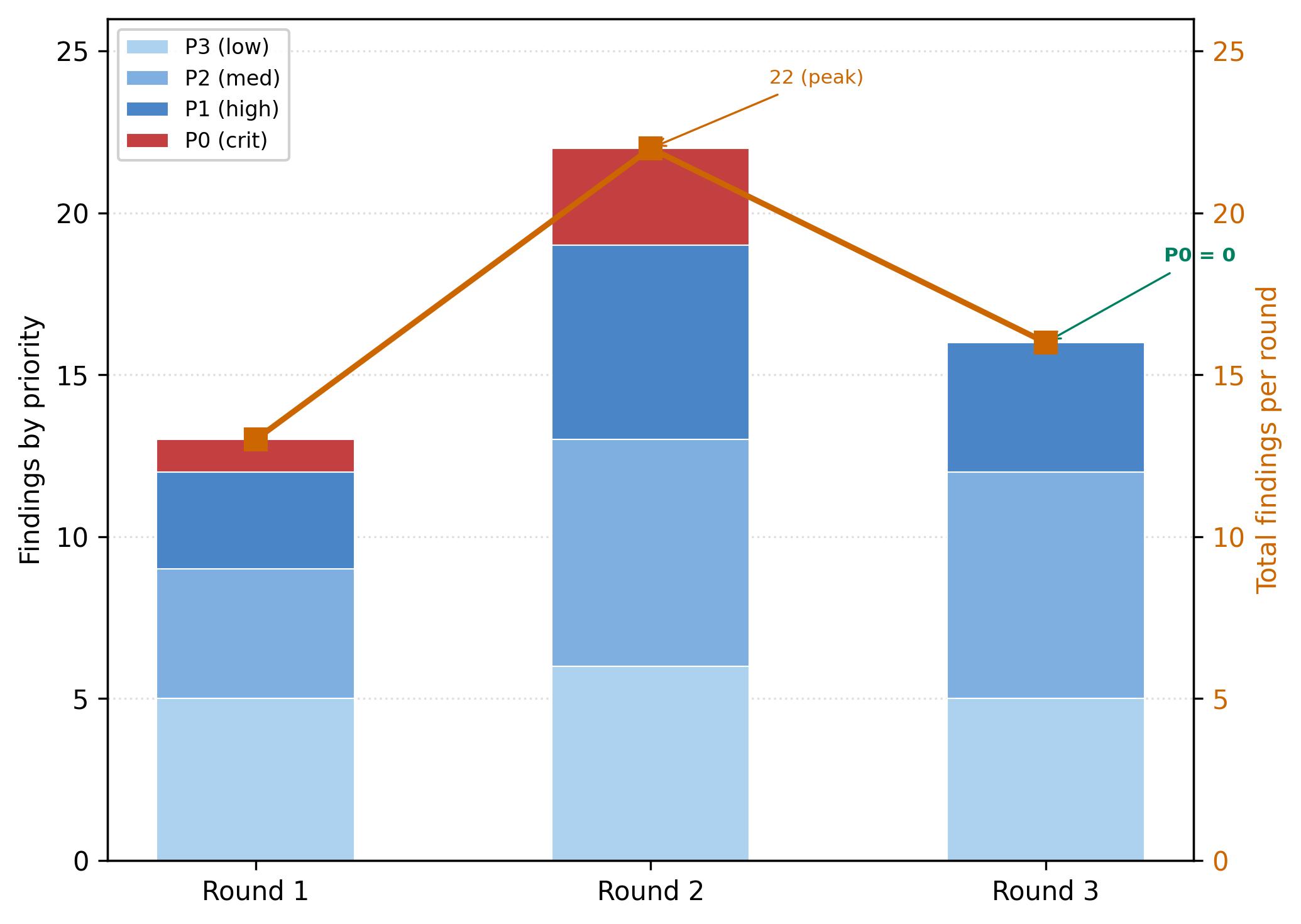}
  \caption{Adversarial code review finding trend across three rounds. P0~(critical)
    findings peaked at~3 in Round~2, exposing second-order vulnerabilities revealed by
    Round~1 fixes, reaching zero in Round~3. The total findings bar chart is shown on the
    secondary axis.}
  \label{fig:p0trend}
\end{figure}

Three rounds of adversarial code review applied a STRIDE threat model~\cite{bib12} with a
two-person review panel not directly involved in system construction. Vulnerability
categories per round were: Round~1 (lock handling, race conditions, and TOCTOU issues);
Round~2 (subprocess management, Jira API, and shell compatibility); and Round~3 (symlink
traversal, clock skew, and resource management). Table~\ref{tab:adversarial} summarizes
the results. Resolution strategy: 42\% verifier improvements, 35\% process-level gating,
and 23\% policy adjustments. Zero false negatives were observed within the author-constructed
injection set. Fig.~\ref{fig:p0trend} shows the P0 trend.

\subsection{Autonomous Security Ticket Family}\label{subsec:autosec}

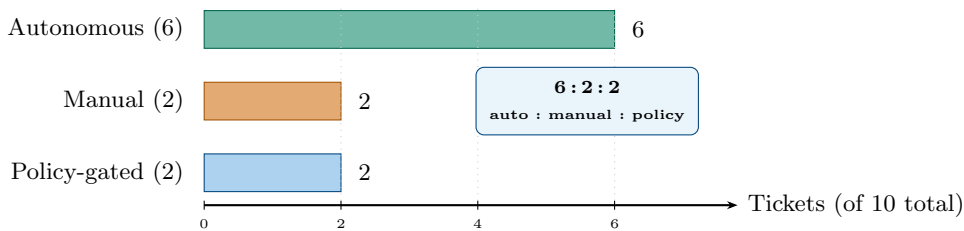
\begin{figure}[htbp]
\centering
\begin{tikzpicture}[x=0.90cm, y=1.0cm]
  %% Three horizontal bars drawn manually — no axis ordering ambiguity
  \def\bh{0.50}   %% bar height
  \def\bsep{0.95} %% vertical separation between bars

  %% Autonomous (6) — green — top bar
  \fill[sngreen!55,draw=sngreen!80!black]
    (0,2*\bsep) rectangle (6,2*\bsep+\bh);
  \node[right=3pt,font=\small] at (6,2*\bsep+0.5*\bh) {6};

  %% Manual (2) — orange — middle bar
  \fill[snorange!55,draw=snorange!80!black]
    (0,\bsep) rectangle (2,\bsep+\bh);
  \node[right=3pt,font=\small] at (2,\bsep+0.5*\bh) {2};

  %% Policy-gated (2) — blue — bottom bar
  \fill[snlblue,draw=snblue]
    (0,0) rectangle (2,\bh);
  \node[right=3pt,font=\small] at (2,0.5*\bh) {2};

  %% y-axis labels
  \node[left=4pt,font=\small,align=right] at (0,2*\bsep+0.5*\bh) {Autonomous (6)};
  \node[left=4pt,font=\small,align=right] at (0,\bsep+0.5*\bh)   {Manual (2)};
  \node[left=4pt,font=\small,align=right] at (0,0.5*\bh)         {Policy-gated (2)};

  %% x-axis
  \draw[-{Stealth[length=4pt]},thick] (0,-0.18) -- (7.8,-0.18)
    node[right,font=\small] {Tickets (of 10 total)};
  \foreach \x in {0,2,4,6}
    \draw (\x,-0.12) -- (\x,-0.24) node[below,font=\tiny] {\x};

  %% vertical grid lines
  \foreach \x in {2,4,6}
    \draw[dotted,gray!40] (\x,-0.12) -- (\x,2*\bsep+\bh+0.08);

  %% 6:2:2 annotation box
  \node[draw=snblue,fill=snlblue!25,rounded corners=3pt,
        font=\footnotesize\bfseries,inner sep=5pt,align=center]
    at (5.6,\bsep+0.5*\bh)
    {6\,:\,2\,:\,2\\[2pt]{\tiny auto : manual : policy}};
\end{tikzpicture}
  \caption{AUTO-SEC ticket outcome distribution. The 6:2:2 ratio (autonomous~: manual~:
    policy) illustrates appropriate boundary enforcement. Pipeline-autonomous tickets
    (SEC-001, -002, -004, -006) required only tool invocation and deterministic output
    parsing; SEC-003 and SEC-005 included human code edits within the autonomously
    dispatched pipeline.}
  \label{fig:autosec}
\end{figure}

Table~\ref{tab:autosec} details the 10~AUTO-SEC ticket outcomes. Six were autonomously
dispatched, monitored, and verified by the pipeline; two required manual completion
(architectural judgment required: security gate channel split and SAST artifact noise
reduction); two were resolved via policy decision. The 6:2:2 split illustrates the boundary
enforcement in practice; the ratio reflects the current policy configuration rather than
a general optimum. For SEC-003 and SEC-005, ``Manual'' in the Tools Invoked column
denotes the remediation action type (a human performed the code change); ``Autonomous
(verified)'' in Completion Mode reflects that the pipeline autonomously dispatched,
monitored gating, and verified ticket closure. Fig.~\ref{fig:autosec} shows the
distribution.

\subsection{Jira Integration Metrics}\label{subsec:jira-metrics}

During the initial 152-run evaluation window, the system generated 38~Jira receipts
affecting 30~unique issues. In the subsequent continuous-operation period
(March~21--31,~2026), 120~additional tickets were transitioned (82 between March~27--28
and 38 on March~26), using 4~transition IDs (11:~To~Do, 21:~In~Progress, 31:~On~Hold,
41:~Done) within the KAN project. Zero duplicate posts were observed across the full
evaluation period, demonstrating three-layer deduplication effectiveness.

\subsection{Alert and Observability Metrics}\label{subsec:alerts}

During the initial evaluation window, the system generated 99~automation alerts across
five categories. Since the transition to continuous per-lane operation, alert categories
have been restructured to align with the 7-lane architecture, with each lane producing its
own alert stream. Total report artifacts have grown from 313 to 795+ as of March~31, 2026,
reflecting the higher cadence of continuous scheduled execution. Alert thresholds remain
adaptive; production configuration continues to benefit from reclassification of stable
dependency-audit findings.

%%=============================================================

\section{Discussion}\label{sec:discuss}
%%=============================================================

\subsection{Lifecycle Automation vs.\ Code Generation}\label{subsec:lifecycle}

Commercial AI-assisted development tools such as GitHub Copilot~\cite{bib1},
Tabnine~\cite{bib2}, and OpenAI Codex~\cite{bib13} emphasize code-oriented assistance.
The present system targets a narrower, bounded problem: orchestrating resolution of a known backlog
under explicit control constraints. The ratio of pipeline-autonomous completions to total
backlog items illustrates the intentional scope boundary. As confidence grows in a task
family, its autonomy level can be promoted via configuration; failures trigger policy
rollback without code changes.

\subsection{Quantitative Safety Argument}\label{subsec:safetyarg}

The safety case rests on four independent quantitative measures: (1)~exception-handler
density of 101~handlers across the automation codebase, a metric reflecting
defensive-coding depth that increased through post-evaluation hardening; (2)~12~centralized
lock mechanisms covering all shared-resource classes, implemented through shared
lock-handler functions and reusable patterns; (3)~checkpoint-based time budgets
guaranteeing liveness; and (4)~full adversarial finding resolution (all 51~closed;
follow-on hardening recommendations issued for 3~P3 items) with zero false negatives across
$n = 51$ injected findings.

The collision-lock protocol and degraded-mode resilience pattern add two additional design
safeguards, although real contention testing and real Jira-outage field validation remain
future work. This forms a partially validated safety argument rather than a formal proof.
Prior work has related component size to software fault behavior~\cite{bib14}, but the
present paper does not infer such a relationship from infrastructure-density percentages
alone. Formal verification of state-machine transitions remains future work.

\subsection{Interpreting the DEGRADED Rate}\label{subsec:degradedrate}

The 79.3\% nightly DEGRADED rate should not be interpreted as failure. All 65~DEGRADED
nightly runs completed successfully. The rating reflects a transparency-first design:
surfacing information for operator awareness rather than suppressing it. This prioritizes
auditability over superficial health metrics, consistent with Section~\ref{sec:compliance}
requirements.

\subsection{Threats to Validity}\label{subsec:threats}

Four categories are identified.

\paragraph{Internal Validity.}
The adversarial review panel consisted of two engineers from the same organization as the
development team, though not directly involved in construction. The vulnerability injection
set was constructed by one of the authors, introducing possible selection bias toward known
vulnerability classes.

\paragraph{External Validity.}
Results derive from a single case study: one codebase, one team, one Jira instance.
Generalizability across different technology stacks, team sizes, regulatory regimes, or
backlog structures has not been established. Specific thresholds and timeout values are
expected to require recalibration in other deployments.

\paragraph{Construct Validity.}
DEGRADED status means ``informational alert present,'' not ``component failure''---a
non-standard usage that may inflate apparent failure rates in cross-study comparisons if
not adjusted.

\paragraph{Statistical Validity.}
Sample sizes are modest ($n = 152$ runs in the initial evaluation window, $n = 51$
adversarial findings) and no control group exists for direct comparison. The 100\%
success-rate claim carries a 95\% CI of $[97.6\%, 100\%]$ (Clopper--Pearson~\cite{bib11}).
No statistical power analysis was conducted before evaluation. Additionally, the collision
lock protocol has been implemented but not yet tested under real multi-agent contention; the
mechanism is architecturally sound but field validation of lock contention rates and race
condition frequency is pending. The degraded-mode protocol has not been exercised under
production Jira downtime. Lane~7 was introduced on March~31, 2026, and has not yet produced
empirical results. The model switch from GPT-5.4 to Claude mid-study complicates
before/after comparisons, though the bounded-autonomy argument is model-agnostic by design.
Finally, the 11-to-7 lane consolidation may shift rather than eliminate complexity.
Additionally, PROMPT versions advanced during the writing period (Lane~2 v2.3.0 to v2.4.0,
Lane~4 v3.5.0 to v3.7.0, Lanes~3/5/6 each advancing one minor version), and some
architectural claims in earlier drafts reflected pre-update behavior; the paper has been
revised to match current deployed versions, but rapid system evolution during documentation
is itself a validity concern. Additionally, the MLT frameworks (MANDATE, LATTICE, TRACE)
have been submitted for peer review but have not yet completed the review process; the
present paper provides independent empirical evidence of their joint operation in a
production system.

\subsection{Ethical Considerations}\label{subsec:ethics}

AI involvement is bounded decision-making within policy rails. The system augments rather
than replaces human judgment: complex decisions remain human-led; routine work is
accelerated. Immutable audit trails and operator override at every gate ensure
accountability.

\subsection{Limitations and Future Work}\label{subsec:futurework}

All 51~adversarial findings were closed within this study's scope: 48 were fully
remediated, while three low-priority P3 items (offline recovery edge cases, health
watchdog timing under clock skew, and eslint/npm-audit false-positive characterization)
were closed with follow-on hardening recommendations deferred to future work. Since the initial evaluation, several future work items have been
partially addressed: autonomous execution has been extended to additional task families
(AUTO-DROP with 442~tickets and AUTO-TECH with 10~tickets), and the degraded-mode protocol
and collision lock mechanism address several P3 hardening recommendations.

Remaining future work includes: MCP error injection testing to simulate Jira failures and
measure queue drain correctness; collision lock contention measurement under parallel
multi-agent execution; Lane~7 empirical results including coverage percentage, patch
accuracy, and false gap rate; degraded-mode field validation under real Jira downtime;
cross-repository federation via MCP multi-project support; formal verification of state
machine transitions; and characterization of static-analysis false-positive rates under the
deployed configuration.

\section{Related Work}\label{sec:related}
%%=============================================================

CI/CD platforms such as Jenkins~\cite{bib15}, GitLab~CI~\cite{bib16}, and GitHub
Actions~\cite{bib17} automate build, test, and deployment. The present work extends this
abstraction stack upward, automating what to build and test rather than the mechanics of
doing so.

Infrastructure-as-Code frameworks such as Terraform~\cite{bib18} and
Ansible~\cite{bib19} express the desired state as declarative specifications. The
dual-representation model presented here parallels IaC: the desired state is declared
locally and synchronized to the authoritative Jira environment.

AI-assisted code synthesis tools~\cite{bib1,bib2,bib13} generate code from context. The
present system differs fundamentally: AI coordinates existing tools and human decisions
rather than generating novel source code.

Issue tracker automation platforms such as Atlassian Automation~\cite{bib20} and
ServiceNow~\cite{bib21} automate routine Jira tasks via trigger-action rules. The system
described here adds semantic domain logic: four-tier fuzzy canonicalization, execution
planning, and formal verification.

Autonomic computing~\cite{bib22} proposes self-managing systems with feedback loops and
recovery. The seven-stage control loop and FMEA-guided failure recovery draw from this
tradition, adapted for deterministic operation. The dependability taxonomy~\cite{bib23}
informs the FMEA construction.

Multi-agent coordination and distributed mutual exclusion are relevant to the collision-lock
protocol described in Section~\ref{subsec:contract}. Lamport~\cite{bib28} established
foundational ordering principles for distributed systems; the Jira Status Contract
implements a pragmatic mutual-exclusion mechanism using Jira status transitions as
distributed locks. The shift from direct HTTP API calls to Model Context Protocol Jira
tools~\cite{bib29} represents a standardized tool-invocation interface for AI-system
integration. The degraded-mode resilience pattern (Section~\ref{subsec:degraded}) draws on
fail-stop processor concepts~\cite{bib30} and event-sourcing replay patterns~\cite{bib31}.
Lane~7's specification-completeness feedback loop addresses specification drift, a
phenomenon Parnas~\cite{bib32} identified as software aging. The self-healing systems
literature~\cite{bib33} provides context for the automated recovery coordination performed
by Lane~5.

The MLT governance stack of Context Engineering~\cite{bib24}, MANDATE~\cite{bib25},
LATTICE~\cite{bib26}, and TRACE~\cite{bib27} provides the theoretical and architectural
foundations instantiated in this case study. Context Engineering~\cite{bib24} formalizes
the human-to-AI input layer; its Reviewer stage is the human-side ingest processor for
Stage~1. MANDATE~\cite{bib25} provides the authorization framework for autonomous agentic
execution. LATTICE~\cite{bib26} defines the confidence-threshold model implemented by the
three canonicalization tiers (${\geq}0.83$ autonomous, $0.50$--$0.83$ human review,
${<}0.50$ halt and re-ingest). TRACE~\cite{bib27} specifies the runtime-evidence
requirements implemented by the paper's log and receipt chain. Each framework was developed
and released individually as a preprint or working paper; this paper reports their joint
empirical deployment in one production system, now operating continuously across 7~lanes
rather than in batch mode.

\section{Conclusion}\label{sec:conclusion}
%%=============================================================

This paper presented a closed-loop autonomous system for software lifecycle management,
structured as a deterministic seven-stage control loop implemented as 7~consolidated
automation lanes operating on continuous scheduled cadences. During the initial 152-run
evaluation window, the system achieved 100\% terminal-state success (95\% CI
[97.6\%, 100\%]) with zero duplicate Jira posts, full adversarial finding closure
(all 51~closed; 48 fully remediated, 3~P3 items closed with deferred hardening
recommendations) with zero false negatives within the injected set, and 6 of 10
autonomous-security tickets resolved through pipeline-autonomous dispatch.

\subsection*{Research Question Responses}

\textbf{RQ1} asked to what extent a deterministic control loop can reduce ad-hoc human
judgment for routine backlog triage. The case study provides observational evidence that
the seven-stage pipeline can automate intake, canonicalization, execution, and
verification for bounded task families (seven families totaling approximately
1,602~rows), reducing routine triage to configuration and exception handling. The system
does not eliminate human judgment across the full lifecycle; complex architectural
decisions, policy choices, and novel task families remain human-led.

\textbf{RQ2} asked where the practical boundary of safe autonomous execution lies and
how it should be enforced. The AUTO-SEC family (10~tickets, 6:2:2 autonomous-to-manual-to-policy
split) demonstrates one observed boundary: tickets with deterministic fix plans and
verifiable outcomes (dependency updates, configuration changes) were suitable for
autonomous dispatch, while tickets requiring architectural judgment or cross-component
coordination required human intervention. The boundary was enforced through
confidence-tier thresholds (${\geq}0.83$ autonomous, $0.50$--$0.83$ human review,
${<}0.50$ halt), policy gates at each pipeline stage, and human override at every
transition point. Whether this boundary generalizes beyond the observed task families
remains an open question.

\textbf{RQ3} asked what safety mechanisms are necessary to achieve a quantitatively
defensible reliability claim without formal proof. The evaluation identified six
mechanisms present in all successful runs: explicit state representation via finite-state
machines, collision locking via the Jira Status Contract, checkpoint-based time budgets,
separated verification (the executor cannot validate its own output), deduplicated
publication through three-layer deduplication, and tamper-evident evidence logging
conforming to the TRACE specification. These mechanisms were individually necessary in
the observed deployment; the paper does not establish that they are jointly sufficient
in all settings.

\subsection*{Broader Implications}

The results suggest that lifecycle automation is a distinct problem from code generation,
requiring control-theoretic structure rather than generative capability. Accountability
in partially autonomous systems comes from explicit control architecture, monitoring,
recovery protocols, and traceable evidence chains rather than from the accuracy of any
single AI component. The bounded-autonomy pattern---where the system's scope is
configured rather than learned, and failures trigger policy rollback rather than
retraining---may be applicable to other domains where deterministic control and
auditability are required.

%%=============================================================

\hfuzz=5pt   %% suppress cosmetic sub-5pt overruns in table cells

%% ------------------------------------------------------------------
%% Table 1: FMEA (5 cols, footnotesize, tabcolsep 4pt)
%% budget = 372 - 5*8 = 332 pt = 11.71 cm
%% cols: 2.7 + 0.8 + 2.9 + 2.9 + 2.41 = 11.71 cm
%% ------------------------------------------------------------------
\setlength{\tabcolsep}{4pt}
\footnotesize
\begin{longtable}{P{2.7cm}C{0.8cm}P{2.9cm}P{2.9cm}P{2.41cm}}
  \caption{Failure Mode and Effects Analysis (19 modes; IEC~60812:2018~\cite{bib8}.
    IV~=~catastrophic, III~=~critical, II~=~marginal, I~=~negligible.)}\label{tab:fmea}\\
  \toprule
  \textbf{Failure Mode} & \textbf{Sev.} & \textbf{Detection Mechanism}
    & \textbf{Recovery Action} & \textbf{Residual Risk} \\
  \midrule
  \endfirsthead
  \multicolumn{5}{l}{\footnotesize\textit{Table~\ref{tab:fmea} continued}}\\
  \toprule
  \textbf{Failure Mode} & \textbf{Sev.} & \textbf{Detection Mechanism}
    & \textbf{Recovery Action} & \textbf{Residual Risk} \\
  \midrule
  \endhead
  \midrule
  \multicolumn{5}{r}{\footnotesize\textit{Continued\ldots}}\\
  \endfoot
  \bottomrule
  \endlastfoot
  Duplicate posting to Jira & I & Receipt log + Jira comment history
    & Abort post; log duplicate event & Negligible \\[2pt]
  State mutation during verify-only & III & \texttt{FAIL\_VERIFY\allowbreak\_MUTATED\allowbreak\_STATE} flag
    & Discard worktree; require manual review & Marginal \\[2pt]
  Process hang beyond timeout & III & Checkpoint model + hard-stop timer
    & USR1 warning; SIGKILL after 30~s & Marginal \\[2pt]
  Jira rate-limit or retryable failure & II & MCP error classification / retryable signal
    & Exponential back-off $\pm$10\% jitter & Negligible \\[2pt]
  Snapshot restore corruption & III & SHA-1 checksum comparison on restore
    & Re-checkout from origin; flag run & Negligible \\[2pt]
  Race condition on receipt & II & 12 centralized lock mechanisms
    & Retry with jittered back-off & Negligible \\[2pt]
  Canonical backlog divergence & III & Symlink check; atomic sync; diff
    & Re-sync Jira from canonical backlog & Marginal \\[2pt]
  Dependency audit cascade alert & I & Alert classification layer
    & Route as informational (non-blocking) & Negligible \\[2pt]
  AI content format error & II & Format re-validation + schema check
    & Reject; escalate for manual review & Negligible \\[2pt]
  Circuit breaker blocks publish & III & Escalation recovery logic
    & Salvage verified items; skip failed & Marginal \\[2pt]
  Offline backlog stale & II & Watermark tracking per run
    & Re-sync on connectivity restoration & Negligible \\[2pt]
  Health watchdog false positive & I & Manual threshold review via UI
    & Operator override via config variable & Negligible \\[2pt]
  Jira connectivity loss during fix & III & \texttt{jira\_connectivity.json} check
    & Degraded mode; queue for Lane~5 replay & Marginal \\[2pt]
  Multi-agent collision on ticket & III & Atomic In~Progress transition
    & Second agent skips; logs collision & Negligible \\[2pt]
  Stale In~Progress (crashed agent) & II & Lane~5 (4h WARNING, 10h CRITICAL)
    & Alert operator; candidate for requeue & Marginal \\[2pt]
  Aging On~Hold ticket & II & Lane~5 monitoring (48h threshold)
    & Alert operator; escalate to digest & Marginal \\[2pt]
  Intra-lane worker collision & II & Adaptive Worker Pool claim queue
    & Worker skips already-claimed ticket & Negligible \\[2pt]
  Worker pool resource exhaustion & II & Time-budget checkpoint (10-min)
    & Hard stop at budget; requeue ticket & Marginal \\[2pt]
  Time-budget exhaustion mid-ticket & II & Checkpoint model timeout
    & Discard worktree; requeue to To~Do & Marginal \\
\end{longtable}
\normalsize
\setlength{\tabcolsep}{6pt}

%% ------------------------------------------------------------------
%% Table 2: System scale metrics (3 cols, tabcolsep 4pt)
%% budget = 372 - 3*8 = 348 pt = 12.27 cm
%% cols: 3.5 + 3.2 + 5.57 = 12.27 cm
%% ------------------------------------------------------------------
\begin{table}[htbp]
  \caption{System scale metrics, automation infrastructure}\label{tab:scale}
  \setlength{\tabcolsep}{4pt}
  \small
  \begin{tabular}{P{3.5cm}P{3.2cm}P{5.57cm}}
    \toprule
    \textbf{Metric} & \textbf{Value} & \textbf{Notes} \\
    \midrule
    Total LOC (automation) & {\small$\sim$12,661 + 6,907 PROMPT} & 23 scripts + 7 prompt files; consolidated from 51 \\
    Total LOC (application) & 1,290,158 & 15,840 source files: JS/TS/CSS, Python, Shell, Config \\
    Backlog rows & $\sim$1,602 & 7 families; CERTREQ removed, AUTO-DROP and AUTO-TECH added \\
    Environment variables & Per lane & Time budgets, ticket caps, MCP rate limits by lane \\
    CLI argparse arguments & 77 & 60 with configurable defaults across 9 scripts \\
    Timeout boundaries & 17 & 45-min standard, 120-min nightly, 10-min progress check \\
    Exception handlers & 101 & Increased from 88 via post-evaluation hardening \\
    Lock mechanisms & 12 & 4 core functions + 8 patterns; 7 per-lane lock files \\
    Jira API methods used & 6 MCP tools & Search, retrieve, create, edit, comment, transition \\
    Fix plans (mapped) & $\sim$120 & Spanning apply, verify, and manual action types \\
    Status outcomes (publication) & 18 & 18 distinct Jira post result codes \\
    Deduplication layers & 3 & Receipt log, Jira history, per-run cache \\
    Source documents ingested & 13 & Across 5 planning surfaces \\
    Test files / test functions & 114 / 1,062 & 19,284 LOC test infrastructure \\
    \bottomrule
  \end{tabular}
\end{table}

%% ------------------------------------------------------------------
%% Table 3: Operational reliability (6 cols, tabcolsep 4pt)
%% budget = 372 - 6*8 = 324 pt = 11.43 cm
%% cols (P types): 2.8+0.6+1.7+1.7+1.1+3.53 = 11.43 cm
%% CI header abbreviated; full citation in caption
%% ------------------------------------------------------------------
\begin{table}[htbp]
  \caption{Operational reliability summary, 152~runs. DEGRADED~= informational
    alerts present; all runs achieved terminal-state success. CI computed via
    Clopper--Pearson exact method~\cite{bib11}.}\label{tab:reliability}
  \setlength{\tabcolsep}{4pt}
  \footnotesize
  \begin{tabular}{P{2.8cm}rP{1.7cm}P{1.7cm}cP{3.53cm}}
    \toprule
    \textbf{Run Type} & \textbf{$n$} & \textbf{CLEAN} & \textbf{DEGRADED}
      & \textbf{Success} & \textbf{95\% CI} \\
    \midrule
    Nightly full sweeps    & 82  & 17 (20.7\%) & 65 (79.3\%) & 100\% & [95.6\%, 100\%] \\
    Executor runs          & 29  & 25 (86.2\%) &  4 (13.8\%) & 100\% & [88.1\%, 100\%] \\
    Remediation runs       & 41  & 33 (80.5\%) &  8 (19.5\%) & 100\% & [91.4\%, 100\%] \\
    \midrule
    \textbf{Overall}       & 152 & 75 (49.3\%) & 77 (50.7\%) & 100\% & [97.6\%, 100\%] \\
    \bottomrule
  \end{tabular}
\end{table}

%% ------------------------------------------------------------------
%% Table 4: Adversarial review (7 cols, tabcolsep 4pt)
%% budget = 372 - 7*8 = 316 pt = 11.15 cm
%% cols: 0.6+4.1+0.6+0.6+0.6+0.6+0.9+2.15 = nope, 7 cols
%% cols: 0.65+4.2+0.6+0.6+0.6+0.6+1.0+2.9 -- that's 8
%% 7 cols: Rnd+Category+P0+P1+P2+P3+Total+Res -- actually 8
%% Use 7 cols: merge P0-P3 into one "Findings" subhead, or keep 8 narrow
%% Keep 8 cols: 0.55+3.5+0.55+0.55+0.55+0.55+0.7+2.1 = 9.05 too tight
%% Better: 7 effective cols by using footnotesize
%% budget footnotesize+tabcolsep4pt: same pt budget but smaller font
%% ------------------------------------------------------------------
\begin{table}[htbp]
  \caption{Adversarial code review findings by round and priority (STRIDE threat
    model~\cite{bib12}; P0~=~critical, P1~=~high, P2~=~medium,
    P3~=~low)}\label{tab:adversarial}
  \setlength{\tabcolsep}{4pt}
  \footnotesize
  \begin{tabular}{cP{3.8cm}cccccc}
    \toprule
    \textbf{Rnd} & \textbf{Injection Focus}
      & \textbf{P0} & \textbf{P1} & \textbf{P2} & \textbf{P3}
      & \textbf{Tot.} & \textbf{Res.} \\
    \midrule
    1 & Lock handling, race conditions, TOCTOU & 1 & 3 & 4 & 5 & 13 & 100\% \\
    2 & Subprocess mgmt, Jira API, shell compat. & 3 & 6 & 7 & 6 & 22 & 100\% \\
    3 & Symlink traversal, clock skew, resources & 0 & 4 & 7 & 5 & 16 & 100\% \\
    \midrule
    All & All rounds & 4 & 13 & 18 & 16 & 51 & 100\% \\
    \bottomrule
  \end{tabular}
\end{table}

%% ------------------------------------------------------------------
%% Table 5: AUTO-SEC tickets (5 cols, footnotesize, tabcolsep 4pt)
%% budget = 332 pt = 11.71 cm
%% cols: 1.6+3.2+3.0+0.8+3.11 = 11.71 cm
%% long \texttt filenames broken with \allowbreak at underscores
%% ------------------------------------------------------------------
%% helper macro: allow breaks at each underscore in \texttt paths
%% helper macro: unused placeholder removed
\begin{table}[htbp]
  \caption{Autonomous security ticket family outcomes. For SEC-003 and
    SEC-005, ``Manual'' denotes remediation action type; ``Autonomous
    (verified)'' denotes pipeline dispatch and verification mode.}\label{tab:autosec}
  \setlength{\tabcolsep}{4pt}
  \footnotesize
  \begin{tabular}{P{1.6cm}P{3.2cm}P{3.0cm}P{0.8cm}P{3.11cm}}
    \toprule
    \textbf{ID} & \textbf{Type} & \textbf{Tools Invoked}
      & \textbf{Status} & \textbf{Completion Mode} \\
    \midrule
    SEC-001 & Runtime dependency remediation (npm audit)
      & \texttt{security-audit.sh} & Done & Autonomous (verified) \\[2pt]
    SEC-002 & Auth token storage hardening
      & \texttt{verify\_bl\_sec\allowbreak\_ticket.py} & Done & Autonomous (verified) \\[2pt]
    SEC-003 & \texttt{dangerouslySet\allowbreak InnerHTML} boot script
      & Human code edit & Done & Autonomous (verified) \\[2pt]
    SEC-004 & innerHTML/XSS surface hardening
      & \texttt{verify\_bl\_sec\allowbreak\_ticket.py} & Done & Autonomous (verified) \\[2pt]
    SEC-005 & \texttt{child\_process} hardening and allowlist
      & Human code edit & Done & Autonomous (verified) \\[2pt]
    SEC-006 & Dev-tool dependency remediation (npm audit moderate)
      & \texttt{security-audit.sh} & Done & Autonomous (verified) \\[2pt]
    SEC-007 & Security gate channel split (runtime vs.\ dev)
      & Manual architectural review & Manual & Manual (architecture) \\[2pt]
    SEC-008 & SAST generated-artifact noise reduction
      & Manual architectural review & Manual & Manual (architecture) \\[2pt]
    SEC-009 & Secret-pattern false-positive cleanup
      & Policy decision & Done & Policy-gated \\[2pt]
    SEC-010 & Lint/SAST auto-fix for debrief panel blockers
      & Policy decision & Done & Policy-gated \\
    \bottomrule
  \end{tabular}
\end{table}
\normalsize
\setlength{\tabcolsep}{6pt}

\backmatter

\section*{Declarations}

\subsection*{Funding}
No external funding was received for this research. The work was conducted as part of
the author's professional responsibilities at The Swift Group, LLC.

\subsection*{Conflict of Interest}
The author is employed by The Swift Group, LLC, which developed and operates the system
described in this paper. The system automates internal software lifecycle processes and
is not a commercial product. The author declares no other competing interests.

\subsection*{Ethics Approval}
Not applicable. This research involved no human subjects. AI involvement is bounded to
supervisory decision-making within deterministic policy rails, as described in
Section~\ref{sec:safety}.

\subsection*{Data Availability}
The automation architecture, pipeline specifications, FMEA analysis, adversarial review
methodology, and evaluation metrics are documented in this paper. The underlying
codebase and Jira project data contain proprietary client information and cannot be
released publicly. The adversarial injection methodology and finding classifications
(Table~\ref{tab:adversarial}) are reported in sufficient detail to enable independent
replication of the threat-model-based review approach. The companion context engineering
dataset and extraction methodology are published as open-access
artifacts~\cite{bib24}.

\subsection*{Author Contributions}
E.\ Calboreanu conceived the architecture, implemented the automation system,
conducted the evaluation, and wrote the manuscript. The adversarial code review was
performed by a two-person panel not directly involved in system construction, as
described in Section~\ref{subsec:adversarial}.

\bibliography{references}

\end{document}